\documentclass{vldb}
\usepackage{graphicx}
\usepackage{balance}
\usepackage{algorithm}
\usepackage{algpseudocode}
\algtext*{EndWhile}
\algtext*{EndIf}
\algtext*{EndFor}
\algtext*{EndProcedure}
\usepackage{enumitem}
\usepackage{verbatim}
\usepackage{pifont}
\usepackage{float}\usepackage{stfloats}
\usepackage{color,soul}
\usepackage[T1]{fontenc}
\usepackage{url}
\usepackage{paralist}
\usepackage{caption}
\usepackage{subcaption}
\usepackage{theorem}
\theoremstyle{definition}
\newtheorem{exmp}{Example}[section]
\newcommand{\Paragraph}[1]{\vspace{-0.02in}\smallskip\noindent{\bf #1.}}
\newdef{definition}{Definition}

\vldbTitle{Coconut: A Scalable Bottom-Up Approach for Building Data Series Indexes}
\vldbAuthors{Haridimos Kondylakis, Niv Dayan, Kostas Zoumpatianos and Themis Palpanas}
\vldbDOI{https://doi.org/10.14778/3184470.3184472}

\begin{document}


\title{Coconut: A Scalable Bottom-Up Approach for\\ Building Data Series Indexes}



%
%
%
%

\numberofauthors{4} 

\author{
%
%
\alignauthor
Haridimos Kondylakis\\
       \affaddr{FORTH-ICS}\\
       \email{\large kondylak@ics.forth.gr}
\alignauthor
Niv Dayan\\
       \affaddr{Harvard University}\\
       \email{\large dayan@seas.harvard.edu}
\and
\alignauthor Kostas Zoumpatianos\\
       \affaddr{Harvard University}\\
       \email{\large kostas@seas.harvard.edu}
\alignauthor Themis Palpanas\\
       \affaddr{Paris Descartes University}\\
       \email{\large themis@mi.parisdescartes.fr}
}

\maketitle
\sloppy
\begin{abstract}

Many modern applications produce massive amounts of data series that need to be analyzed, requiring efficient similarity search operations.
However, the state-of-the-art data series indexes that are used for this purpose do not scale well for massive datasets in terms of performance, or storage costs.
We pinpoint the problem to the fact that existing summarizations of data series used for indexing cannot be sorted while keeping similar data series close to each other in the sorted order.
This leads to two design problems.
First, traditional bulk-loading algorithms based on sorting cannot be used.
Instead, index construction takes place through slow top-down insertions, which create a non-contiguous index that results in many random I/Os.
Second, data series cannot be sorted and split across nodes evenly based on their median value; thus, most leaf nodes are in practice nearly empty.
This further slows down query speed and amplifies storage costs.
%
To address these problems, we present Coconut.
The first innovation in Coconut is an inverted, sortable data series summarization that organizes data series based on a z-order curve, keeping similar series close to each other in the sorted order.
As a result, Coconut is able to use bulk-loading techniques that rely on sorting to quickly build a contiguous index using large sequential disk I/Os.
We then explore prefix-based and median-based splitting policies for bottom-up bulk-loading, showing that median-based splitting outperforms the state of the art, ensuring that all nodes are densely populated.
Overall, we show analytically and empirically that Coconut dominates the state-of-the-art data series indexes in terms of construction speed, query speed, and storage costs. 
\\\\\\\\

\end{abstract}

\section{Introduction}
Many scientific and business applications today produce massive amounts of data series\footnote{Informally, a \emph{data series}, or \emph{data sequence}, is an ordered sequence of data points. If the dimension that imposes the ordering of the sequence is time then we talk about \emph{time series}, though a series can also be defined over other measures (e.g., angle in radial profiles in astronomy, mass in mass spectroscopy, position in genome sequences, etc.). For the rest of this paper, we are going to use the terms \emph{data series} and \emph{sequence} interchangeably.} and need to analyze them, requiring the efficient execution of  similarity search, or nearest neighbor operations.
Example applications range across the domains of audio~\cite{KashinoSM99}, images~\cite{YeK09}, finance~\cite{Shasha99}, telecommunications~\cite{humanbehaviorpatterns,DBLP:conf/edbt/MirylenkaCPPM16}, environmental monitoring~\cite{zzz}, scientific data~\cite{HuijseEPPZ14,url:adhd,VALMOD}, and others.

As the price of digital storage continues to plummet, the volume of data series collections grows,  driving the need for the development of efficient sequence management systems~\cite{DBLP:journals/sigmod/Palpanas15,DBLP:conf/ieeehpcs/Palpanas17,KostasThemisTalkICDE}. 
For the specific problem of sequence similarity search, searching for a nearest neighbor by traversing the entire dataset for every query quickly becomes intractable as the dataset size increases.
Consequently, multiple data series indexing techniques have been proposed over the past decade to organize data series based on similarity~\cite{DBLP:conf/sofsem/Palpanas16}. 
The state-of-the-art approach is to index data series based on smaller summarizations that approximate the distances among data series.
This enables pruning large parts of the dataset that are guaranteed to not contain the nearest neighbor, and thereby these indexes significantly improve query speed.

Large data series collections and indexes that span hundreds of gigabytes to terabytes~\cite{url:adhd,url:sds,DBLP:journals/pvldb/PelkonenFCHMTV15} must reside in secondary storage devices for cost-effectiveness.
The problem with such devices is that they incur slow I/Os.
To facilitate nearest neighbor search in such big data applications, it is crucial to be able to construct and query a data series index as fast and cost-effectively as possible\footnote{Note that recent state-of-the-art serial scan algorithms~\cite{rakthanmanon2012searching,DBLP:conf/icdm/MueenHE14} are only efficient for scenarios that involve nearest neighbor operations of a short query subsequence against a very long data series. On the contrary, in this work, we are interested in finding similarities in very large collections of short sequences.}.


\Paragraph{The Problem: Unsortable Summarizations}
In this paper, we show that the state-of-the-art data series indexes exhibit performance and storage overheads that hinder their ability to scale for massive datasets.
We pinpoint the problem to the fact that the summarizations, used as keys by data series indexes, are unsortable.
Existing summarizations partition and tokenize data series into multiple (independent) segments that are laid out in the summarized representation based on their original order within the data series; thus, sorting based on these summarizations places together data series that are similar in terms of their beginning (i.e., the first segment), yet arbitrarily far in terms of the rest of the values\footnote{This is analogous to sorting points in a multi-dimensional space based on one dimension.}. Hence, summarizations cannot be sorted while keeping similar data series next to each other in the sorted order. This leads to the following problems. 



First, traditional bulk-loading algorithms that rely on sorting cannot be used.
Instead, state-of-the-art data series indexes perform bulk-loading through top-down insertions and splitting nodes as they fill up~\cite{DBLP:conf/sofsem/Palpanas16,isax2plus,ZoumpatianosIP16}. This approach leads to many small random I/Os to secondary storage that slow down construction speed. Moreover, the resulting nodes (after many splits) are non-contiguous in storage, meaning that querying involves many slow random I/Os.



The second problem is that it is not possible to sort and thereby split data series evenly across nodes (i.e., using the median value as a splitting point). Instead, state-of-the-art data series indexes divide data series across nodes based on common prefixes across all segments. As a result, it is impossible for entries that do not share a common prefix in one or more of the segments to reside in the same node. We show that this leads to most nodes being nearly empty (i.e., their  fill-factor is low, which translates to an increased number of leaves). This slows down query speed and amplifies storage costs.

\Paragraph{Our Solution: Sortable Summarizations and Coconut}
To address these problems, we show how to transform existing data series summarizations into \textit{sortable summarizations}.
The core idea is interweaving the bits that represent the different segments, such that the more significant bits across all segments precede all less significant bits.
As a result, we describe the first technique for sorting data series based on their summarizations: the series are positioned on a z-order curve~\cite{morton1966}, in a way that similar data series are close to each other.

Moreover, we show that indexing based on sortable summarizations has the same ability as existing summarizations to prune parts of the index that do not contain the nearest neighbor, while it offers two additional benefits: it enables (i) efficiently bulk-loading the index, and (ii) packing data series more densely into nodes. Furthermore, we show that using sortable summarizations enables data series indexes to leverage a wide range of indexing infrastructure.


We further introduce the \textbf{Co}mpact and \textbf{Con}tiguous Seq\textbf{u}ence Infras\textbf{t}ructure (Coconut).
Coconut is a novel data series indexing infrastructure that organizes data series based on \emph{sortable} summarizations.
It uses bulk-loading techniques to quickly build a contiguous index, thereby eliminating random I/O during construction \emph{and} querying.
Furthermore, it is able to split data series across nodes by sorting them and using the median value as a splitting point, leading to data series being packed more densely into leaf nodes (i.e., at least half full).

In order to study the design space and isolate the impact of the different design decisions, we introduce two variants: Coconut-Trie and Coconut-Tree, which split data series across nodes based on common prefixes and median values, respectively.
We show that Coconut-Trie dominates the state-of-the-art in terms of query speed because it creates contiguous leaves. We further show that Coconut-Tree dominates Coconut-Trie and the state-of-the-art in terms of construction speed, query speed \emph{and} storage overheads because it creates a contiguous, balanced index that is also densely populated. Overall, we show across a wide range of workloads and datasets that Coconut-Tree improves both construction speed and storage overheads by one order of magnitude and query speed by two orders of magnitude relative to the state-of-the-art.

Our contributions are summarized as follows.
\begin{compactitem} 
	\item We show that existing data series summarizations cannot be sorted in a straightforward way. Consequently, state-of-the-art data series indexes cannot efficiently bulk-load and pack data densely into nodes, leading to large storage overheads and performance bottlenecks for both index construction and query answering, when dealing with very large data series collections.
	\item We introduce a \emph{sortable} data series summarization that keeps similar data series close to each other in the sorted order, and preserves the same pruning power as existing summarizations. We show how sortability enables new design choices for data series indexes,  thereby opening up infrastructure possibilities that were not possible in the past.
    \item We introduce Coconut-Trie that exploits sortable summarizations for prefix-based bulk-loading of existing state-of-the-art indexes, leading to improvements at querying time performance. 
	\item We present Coconut-Tree, which employs median-based bulk-loading to quickly build the index and to restrict space-amplification, by enabling entries that do not share a common prefix to be in the same node.	
	\item Our experimental evaluation with a variety of synthetic and real datasets demonstrates that Coconut-Tree strictly dominates existing state-of-the-art indexes in terms of both construction speed and storage overheads by one order of magnitude, and query speed by two orders of magnitude.
\end{compactitem}

\section{Preliminaries and Related Work} 
\label{sec:background}

\Paragraph{Data Series}
Measuring data that fluctuate over a dimension is a very frequent scenario in a large variety of domains and applications.
Such data are commonly called data series or sequences. The dimension over which they fluctuate can range from time, angle or position to any other dimension. They can be measured at either fixed or variable intervals.
\begin{definition}
Formally, a data series $s=\{r_1, ..., r_n\}$ is defined as an ordered set of recordings, where each $r_i = <p_i, v_i>$ describes a value $v_i$ corresponding to a position $p_i$.
\end{definition}


\Paragraph{Nearest Neighbor Search}
Analysts perform a wide range of data mining tasks on data series including clustering~\cite{keogh1998,liao2005,rodrigues2008,rakthanmanon2011}, classification and deviation detection~\cite{Shieh2009,Shandola2009}, frequent pattern mining~\cite{DBLP:journals/datamine/MueenKZCWS11,DBLP:journals/tkdd/GrabockaSS16}, and more.
Existing algorithms for executing these tasks rely on performing fast similarity search across the different data series.
Thus, efficiently processing nearest neighbor (NN) queries is crucial for speeding up the aforementioned tasks.
NN queries are formally defined as follows.

\begin{definition}
Given a set of data series $\bold{S} \subseteq \mathcal{S}$, where $\mathcal{S}$ is the set of all possible data series, a query data series $s_q \in \mathcal{S}$ and a distance function $d(\bullet, \bullet) : \mathcal{S} \times \mathcal{S} \rightarrow \mathbb{R}$, a nearest neighbor query is defined as:
$$
nn_{d(\bullet, \bullet)}(s_q, \bold{S}) = s_i \in \bold{S} : d(s_i, s_q) \leq d(s_j, s_q) \forall s_j \neq s_i \in \bold{S}.
$$
\end{definition}

Common distance metrics for comparing data series include Euclidean Distance (ED) and dynamic time warping (DTW).
While DTW is better for most data mining tasks, the error rate using ED converges to that of DTW as the dataset size grows~\cite{Ratanamahatana2005, Xiaopeng2006, Shieh2008}.
Therefore, data series indexes for massive datasets use ED as a distance metric~\cite{Shieh2008,Shieh2009,Zoumpatianos2014,Zoumpatianos2015rinse,ZoumpatianosIP16}, though simple modifications can be applied to make them compatible with DTW~\cite{Shieh2008, Kate2016}. Euclidean distance is computed as the sum of distances between pairs of aligned points in sequences of the same length, where normalizing the sequences for alignment and length is a pre-processing step~\cite{Shieh2008,Shieh2009,Zoumpatianos2014,Zoumpatianos2015rinse,ZoumpatianosIP16}.
In all cases, data are z-normalized by subtracting the mean and dividing by the standard deviation (note that minimizing ED on z-normalized data is equivalent to maximizing their Pearson's correlation coefficient~\cite{MueenNL10}).



{\color{red}
}

\begin{figure}[tb]
	\center
	\includegraphics[width=0.9\columnwidth]{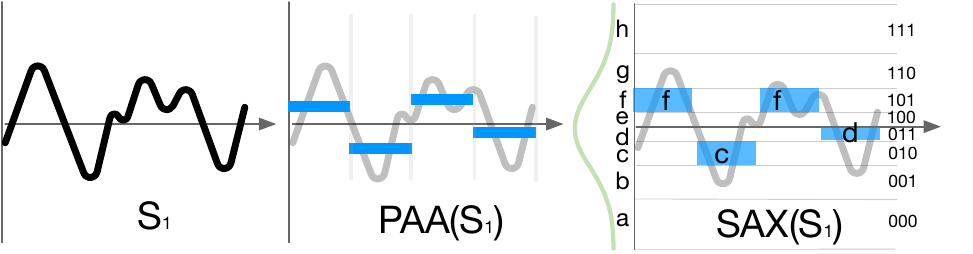}
	\caption{Example PAA and SAX summarizations}\label{fig:sax}
\end{figure}

\Paragraph{Summarizing and Filtering Data Series}
The brute-force approach for evaluating nearest neighbor queries is by performing a sequential pass over the complete dataset.
However, as data series collections grow to terabytes \cite{url:adhd,url:sds,DBLP:journals/pvldb/PelkonenFCHMTV15}, scanning the complete dataset becomes  performance bottleneck taking hours or more to complete. This is especially problematic in exploratory search scenarios, where batch execution of queries is impossible because the next query depends on the results of previous queries. 

To mitigate this problem, various dimensionality reduction techniques have been proposed to transform data series into summarizations that enable approximating and lower bounding the distance between any two data series.  Examples include generic Discrete Fourier Transforms (DFT)~\cite{Agrawal1993,DBLP:conf/sigmod/FaloutsosRM94,Rafiei1997,Rafiei1998}, Piecewise Linear Approximation (PLA)~\cite{Keogh1997}, Singular Value Decomposition (SVD)~\cite{Korn97,RaviKanth1998}, Discrete Haar Wavelet Transforms (DHWT)~\cite{ChanF99, DBLP:conf/kdd/KashyapK11}, Piecewise Constant Approximation (PCA), and Adaptive Piecewise Constant Approximation (APCA)~\cite{Chakrabarti2002}, as well as data series specific techniques such as Piecewise Aggregate Approximation (PAA)~\cite{Keogh2000}, Symbolic Aggregate approXimation (SAX)~\cite{Lin2003} and the indexable Symbolic Aggregate approXimation ($i$SAX)~\cite{Shieh2008,Camerra2010}.
These smaller summarizations can be scanned and filtered~\cite{DBLP:conf/kdd/KashyapK11,Li1996} or indexed and pruned~\cite{Guttman1984, ZoumpatianosIP16, Assent2008, Wang2013, Shieh2008,Shieh2009, Zoumpatianos2014,Zoumpatianos2015rinse,ZoumpatianosIP16,DBLP:conf/icdm/YagoubiAMP17,ulisse} to avoid accessing parts of the data that do not contain the nearest neighbor. 

Our work follows the same high-level idea of indexing the data series based on a smaller summarization to enable pruning, though our work is the first to use sortable summarizations to facilitate index construction. In all previous work, the index is constructed through top-down insertions that lead to many slow random I/Os and to a sparsely populated, non-contiguous and unbalanced index. Our work is the first to use a bottom-up bulk-loading algorithm and median-based splitting to efficiently build a contiguous, balanced, and densely populated index. Note that our infrastructure can be used in conjunction with any summarization that represents a sequence as a multi-dimensional point, and so it is compatible with all main-stream summarizations~\cite{Agrawal1993,DBLP:conf/sigmod/FaloutsosRM94,Rafiei1997,Rafiei1998, Keogh1997, Korn97,RaviKanth1998, ChanF99, DBLP:conf/kdd/KashyapK11, Chakrabarti2002, Shieh2008,Camerra2010}.





\Paragraph{Data Series Indexing with SAX}
We now discuss the state-of-the-art in data series indexing.
We concentrate on SAX summarizations \cite{Lin2003, Shieh2008}, which have been shown to outperform other summarizations in terms of pruning power using the same amount of bytes~\cite{ZoumpatianosLPG15}.
We illustrate the construction of a SAX summarization in Figure \ref{fig:sax}.

SAX first partitions the data series in equal-sized segments, and for each segment it computes its average value.
This is essentially a PAA summarization, and can be seen in Figure~\ref{fig:sax}(middle).
In a second step, it discretizes the value space by partitioning it in regions, whose size follows the normal distribution.
As a result, we have more regions corresponding to values close to 0, and less regions for the more extreme values (this leads to an approximately equal distribution of the raw data series values across the regions, since extreme values are less frequent than values close to 0 for z-normalized series).
A bit-code (or a symbol) is then assigned to every region.
The data series is then summarized by the sequence of symbols of the regions in which each PAA value falls.

In the example in Figure~\ref{fig:sax}, the data series $S_1$ becomes ``fcfd''.
This lossy representation requires much less space (typically in the order of 1\%) and reduces the number of dimensions from the number of points in the original series to the number of segments in the summarization (four in Figure~\ref{fig:sax}).

Data series indexes based on SAX rely on a multi-resolution indexable SAX representation (iSAX)~\cite{Shieh2008,Shieh2009} whereby every node corresponds to a common SAX prefix from across all segments. When a node fills up, the segment whose next unprefixed digit divides the resident data series most is selected for splitting the data series across two new nodes. $i$SAX 2.0~\cite{Camerra2010} and $i$SAX 2+~\cite{isax2plus} are variants that improve construction speed by storing all internal nodes in main memory and buffering access to leaf nodes.
ADS represents the state-of-the-art and builds on these ideas by constructing an index based on the summarizations and later incorporating the raw data series into the index adaptively during query processing. These indexes all share three performance problems. Firstly, if main memory is small relative ot the raw data size, they incur many random I/Os due to swapping and early flushing of buffers. This significantly elongates construction time for massive datasets.
Secondly, the resulting leaf nodes after many splits are non-contiguous in secondary storage and therefore require many slow random I/Os to query. Thirdly, data series that do not share common prefixes cannot reside in the same node, and so the leaf nodes in these indexes are in practice sparsely populated. This leads to significant storage overheads and slows down queries as they must traverse a greater physical area to access the same data.




In contrast to these works, Coconut makes  SAX summarizations sortable (rather than multi-resolution) by inverting their bits. As a result, Coconut is able to use bulk-loading techniques that rely on sorting to quickly build a contiguous index using less I/Os and issuing them sequentially. Moreover, Coconut is able to use median-based splitting to ensure that the index is balanced and densely populated (i.e., all nodes are at least half full).



\section{Problem: Unsortable Summarizations}
\label{sec:problem}

\begin{figure}[tb]
    \centering
    \includegraphics[width=1\columnwidth]{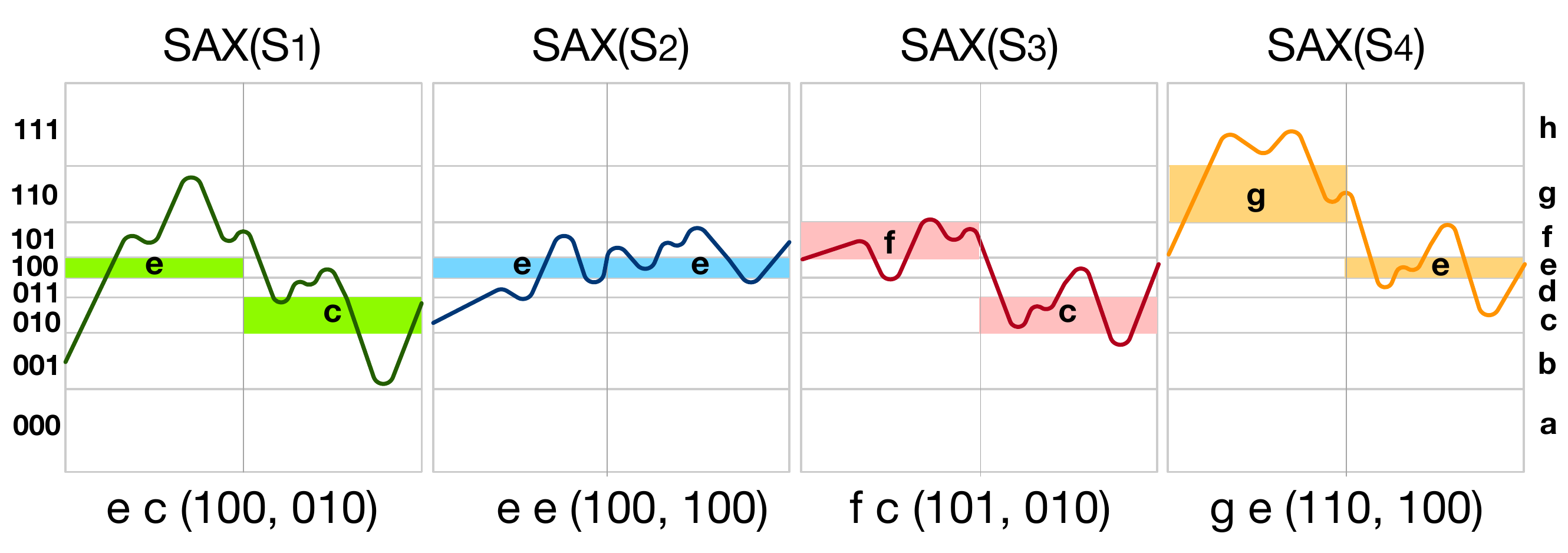}
    \caption{Sorting $i$SAX summarizations}
    \label{fig:sorting}
\end{figure}

In this section, we describe why existing data series summarizations are not sortable, and we discuss the implications on index design, performance, and storage overheads. 

\Paragraph{Sorting summarizations}
Figure~\ref{fig:sorting} on the right gives an example of sorting data series based on SAX summarizations.
There are four different data series with corresponding 2-character SAX words\footnote{Note that SAX words are typically longer to enable more precision; we use 2-character SAX words in this example for ease of exposition.}: $S_1 = ec$, $S_2 = ee$, $S_3 = fc$, and $S_4 = ge$.
Observe that $S_1$ is most similar to $S_3$, while $S_2$ is most similar to $S_4$ (apart from small differences in the first segments).
Sorting these summarizations lexicographically gives the order $S_1, S_2, S_3, S_4$: the data series that are most similar to each other are \emph{not} placed next to each other in the sorted order.
The reason is that existing summarizations lay out the segment representations sequentially, one by one.
Sorting based on such a representation places next to each other data series that are similar in terms of their first segment, yet arbitrarily dissimilar in terms of the rest of the segments.
As a result, an index that is built by sorting data series based on existing summarizations degenerates to scanning the full dataset for each query and defeats the point of having an index.

It is important to note that even though we use SAX, the same observations hold for all other main-stream summarizations (discussed in Section~\ref{sec:background}).
This is because they all represent data series as multi-dimensional points.
As a result, they still suffer from the problem of poor lexicographical ordering, where sorting is based on arbitrarily ordering dimensions.
SAX was chosen in our work, since it has been shown to outperform other approaches in terms of quality~\cite{ZoumpatianosLPG15} and index performance~\cite{Camerra2010,isax2plus,Zoumpatianos2014}.

We next discuss how existing data series indexes overcome the inability to sort summarizations, and we analyze the impact on performance and storage overheads.




\begin{figure}[tb]
	\centering
  \includegraphics[width=0.9\columnwidth]{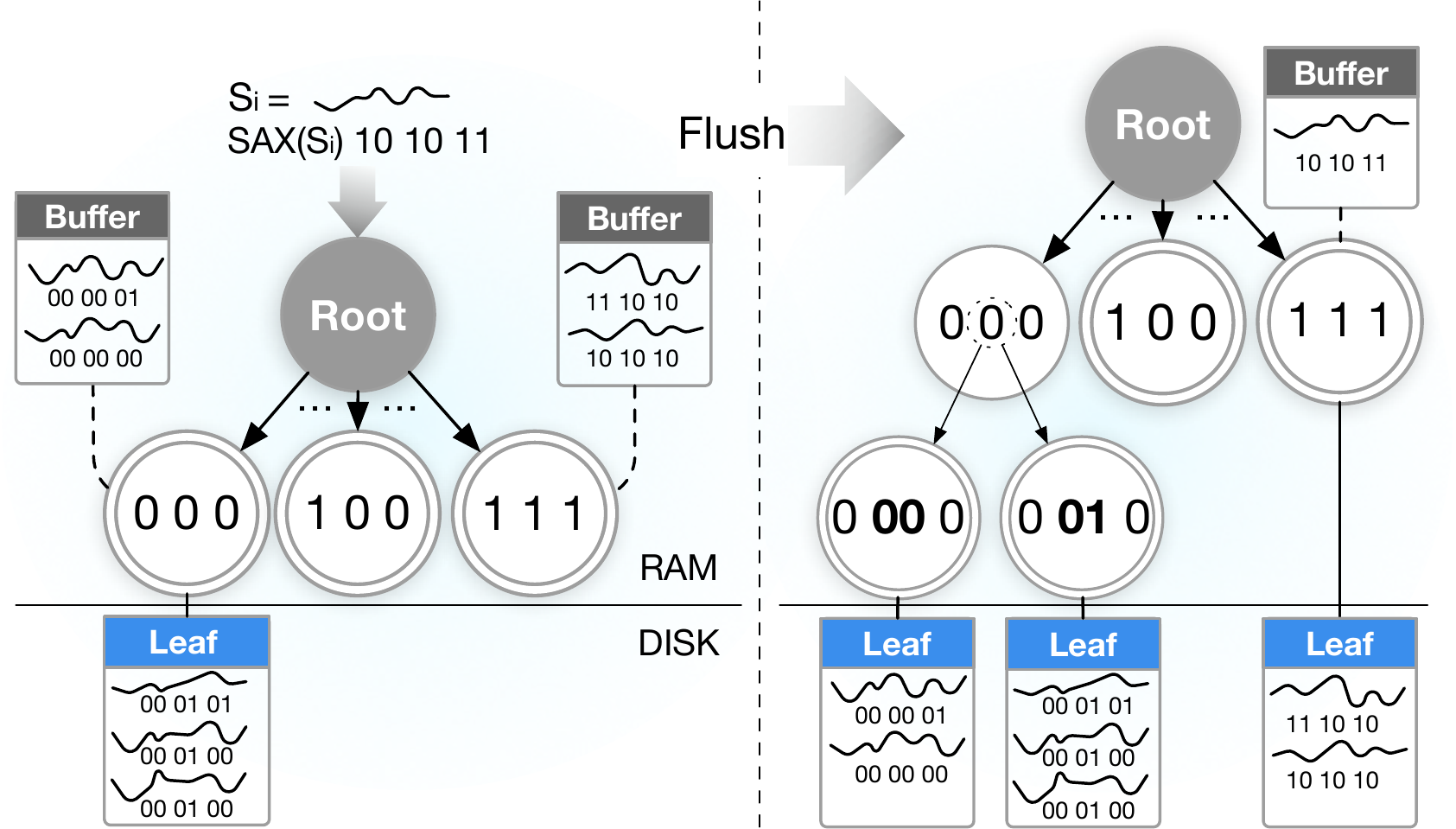}
  \caption{Indexing using $i$SAX 2.0}
  \label{fig:isaxindexing}
\end{figure}
\begin{table}[ht]
\begin{center}
    \begin{tabular}{ | c | l | }
        \hline
        \textbf{Term} & \textbf{Definition} \\
        \hline
        N & Total number of data series \\  \hline
        B & Number of data series that fit into one disk block \\  \hline
        M & Number of data series that fit into main memory  \\
        \hline
    \end{tabular}
    \caption{Table of terms}
    \label{tab:terms}
\end{center}
\end{table}

\subsection{Index Construction}

The standard approach for bulk-loading a database index (e.g., a B-Tree) relies on external sorting.
This approach cannot be used with existing data series summarizations, because they are not sortable.
Instead, state-of-the-art data series indexes perform top-down insertions~\cite{isax2plus,Wang2013,ZoumpatianosIP16}.
Here we analyze and compare their implications on performance and storage overheads. We analyze them in the disk access model \cite{Aggarwal1988}, which measures the runtime of an algorithm in terms of disk blocks transferred between main memory and secondary storage. The terms we use are in Table \ref{tab:terms}.

\Paragraph{Top-Down Insertions}
Building a data series index usually involves top-down insertions: data series are inserted through the root node and they trickle down to the appropriate leaf node~\cite{Shieh2008,Shieh2009}.
This process is illustrated in Figure~\ref{fig:isaxindexing} for the $i$SAX 2.0 index.
Data series are summarized using $i$SAX, and inserted into the tree.
The new series to be inserted, $S_i$, is translated to the $i$SAX word (10 10 11).
At the first level of the tree only the first bit of every segment is considered.
As a result $S_i$ should follow the (1 1 1) sub-tree.
In order to optimize insertion cost, inserts are buffered in main memory.
In our example, all buffers are full (Figure~\ref{fig:isaxindexing}(left)), and have to be processed.
During this operation, when a node runs out of capacity, it creates two new children by increasing the number of bits used to represent one of the segments (we discuss this in detail in Section~\ref{sec:splitting}), and divides the data series between them.
In Figure~\ref{fig:isaxindexing}, node (0 0 0) splits into (0 00 0) and (0 01 0).
This algorithm produces an unbalanced tree index.
During construction, every leaf node is allocated with free space to be able to absorb new insertions.
Every time a leaf has to split, it needs to be re-read from disk and re-written, resulting in multiple reads and writes of the same data.
This can be seen in Figure~\ref{fig:isaxindexing}, where while flushing the buffers, the left-most leaf is re-read and merged with buffered data.
Moreover, new children are allocated wherever there is space on disk, meaning that the original node and the new nodes are not contiguous.
These drawbacks would be essentially eliminated if data were bulk loaded in a bottom up fashion.


Since data series summarizations are much smaller than the raw data of the original data series, the index's internal nodes for most applications fit in main memory \cite{Camerra2010, isax2plus}. Hence, every top-down insertion involves two I/Os: one to read the appropriate leaf and one to update it.
The index construction cost is therefore $O(N)$ I/Os.

As we have seen, buffering of insertions in main memory is used by data series indexes, such as $i$SAX 2.0~\cite{Camerra2010}, in order to amortize the cost of I/O across multiple insertions.
However, buffering does not reduce the  worst-case construction cost of $O(N)$ I/Os as it is easy to construct workloads that involve only cache misses, even when main memory is plentiful.

For uniformly randomly distributed insertions, buffering reduces construction cost to on average $O(N - M + \frac{M}{B})$ I/Os, where $N - M$ is the expected probabilistic number of cache misses\footnote{A cache miss occurs with a probability of $1 - \frac{M}{N}$ and there are $N$ insertions, so the expected number of cache misses is the product of these terms.} and $\frac{M}{B}$ I/Os are needed at the end to write the cached leafs to secondary storage\footnote{There are $M$ buffered data series in main memory, and they occupy $\frac{M}{B}$ disk blocks.   }.
For large datasets, note that $M$ is typically two orders of magnitude lower than $N$ because main memory is two orders of magnitude more expensive than secondary storage, and so buffering is insufficient for ensuring fast index construction speed for massive datasets.



\Paragraph{Bottom-up Bulk-Loading Using External Sorting} Building a database index through external sorting comprises two phases: partitioning and merging.
The partitioning phase involves scanning the raw file in chunks that fit in main memory, sorting each chunk in main memory, and flushing it to secondary storage as a sorted partition. This amounts to two passes over the data. The merging phase involves merge-sorting all the different partitions into one contiguous sorted order, using one input buffer for each partition and one output buffer for the resulting sorted order.
Once the data is ordered, we build the index bottom-up.
Thus, the merging phase amounts to two additional passes over the data, and so external sorting involves overall four passes over the data.
This amounts to $O(N/B)$ I/Os\footnote{In fact this condition only holds as long as $M > \sqrt{N}$ \cite{DBLP:books/daglib/0011128}.
Since main memory is approximately two orders of magnitude more expensive than secondary storage, this condition holds in practice for massive datasets.}.


\Paragraph{Comparison} The analysis in the disk access model above shows that external sorting dominates top-down insertions in terms of worst-case index construction cost because we only need to do a few passes amounting to $O(N/B)$ I/Os rather than $O(N)$ random I/Os. Since a disk block $B$ is typically large relative to data elements, this amounts to a 1-2 order of magnitude difference in construction speed.

External sorting has two additional performance advantages for subsequent query processing. Firstly, the sorted order can be written contiguously in secondary storage, meaning that queries can traverse leaves using large sequential I/Os rather than small random I/Os. Secondly, it is possible to pack data series as compactly as possible in nodes rather than leaving free space for future insertions. This saves storage costs and speeds up queries by reducing the physical space that a query must traverse by a factor of 2.

Overall, external sorting dominates top-down insertions in terms of both construction and query speed. The problem is that existing data series indexes cannot use external sorting as they cannot sort the data based on existing data series summarizations.

\subsection{Splitting Nodes} \label{sec:splitting}

Database indexes such as B-trees split nodes when they run out of capacity using the median value as a splitting point, whereas data series indexes use prefix-based splitting. We now describe these methods in detail and analyze their implications on performance and storage overheads. We again use the disk access model \cite{Aggarwal1988} to quantify storage overheads in terms of disk blocks.


\Paragraph{Prefix-Based Splitting} In state-of-the-art data series indexes, every node is uniquely identified by one prefix for every segment of the SAX representation, and all elements in the node or its subtrees have matching prefixes for all segments. When a leaf node runs out of capacity, we scan the summarizations and identify the segment whose next unprefixed bit divides the elements most. We create two new children nodes and divide the elements among them based on the value of this bit. The problem is that data is not guaranteed to be unevenly distributed across the nodes. In the worst-case, every node split divides the entries such that one moves to one of the new nodes and the rest move to the other, meaning that the index is unbalanced, most nodes contain only 1 entry, and so storage consumption is $O(N)$ disk blocks.


\Paragraph{Median-Based Splitting} Splitting a node using the median value involves sorting the data elements to identify the median, moving all elements to the right of this mid-point into a new node, and adding a pointer from the parent to the new node to ensure the index remains balanced. This approach ensures that every node is at least half full. As a result, the amount of storage space needed is at most double the size of the actual data. This amounts to $O(N/B)$ blocks. 


\Paragraph{Comparison} Prefix-based splitting results in an unbalanced index amplifies worst-case storage overheads relative to median-based splitting by a factor of $B$. Since exact query answering time is proportional to the number of leaf nodes in the index, it amplifies it  by the same factor. Overall, median-based splitting dominates prefix-based splitting, but we cannot use it in the context of data series indexing because existing summarizations are not sortable.

\section{Coconut}
\label{sec:bottomupindexconstruction}

In this section, we present Coconut in detail. Coconut is a novel data series indexing infrastructure that organizes data series based on sortable summarizations. As a result, Coconut indexes are able to use bulk-loading techniques based on sorting to efficiently build a contiguous index. Furthermore, they are able to divide data series among nodes based on sorting and median values to ensure that the index is balanced and that all nodes are densely populated.

In Section~\ref{sec:sortable}, we first show how to make existing summarizations sortable using a simple algorithm that interleaves the bits in a summarization such that all more significant bits from across all segments precede all less significant bits.
In Sections~\ref{sec:trie} and~\ref{sec:tree}, we introduce Coconut-Trie and Coconut-Tree, respectively.
These data structures allow us to isolate and study the impact of the properties of contiguity and compactness on query and storage overheads.

\newpage
\subsection{Sortable Summarizations}
\label{sec:sortable}

Conceptually, sorting data is an operation that involves recursively dividing data entries based on the most significant bit into a hierarchy of sets, and then laying out the elements in the hierarchy in a depth-first order.
We observe that in existing data series summarizations, every subsequent bit in a segment contains a decreasing amount of information about the location of the data point that it represents and simply increases the degree of precision.
Therefore, interleaving the bits for the different segments such that all significant bits precede all less significant bits makes sorting place data series that are similar across all segments next to each other.
\begin{figure}[tb]
	\centering
	\includegraphics[width=0.9\columnwidth, angle=0]{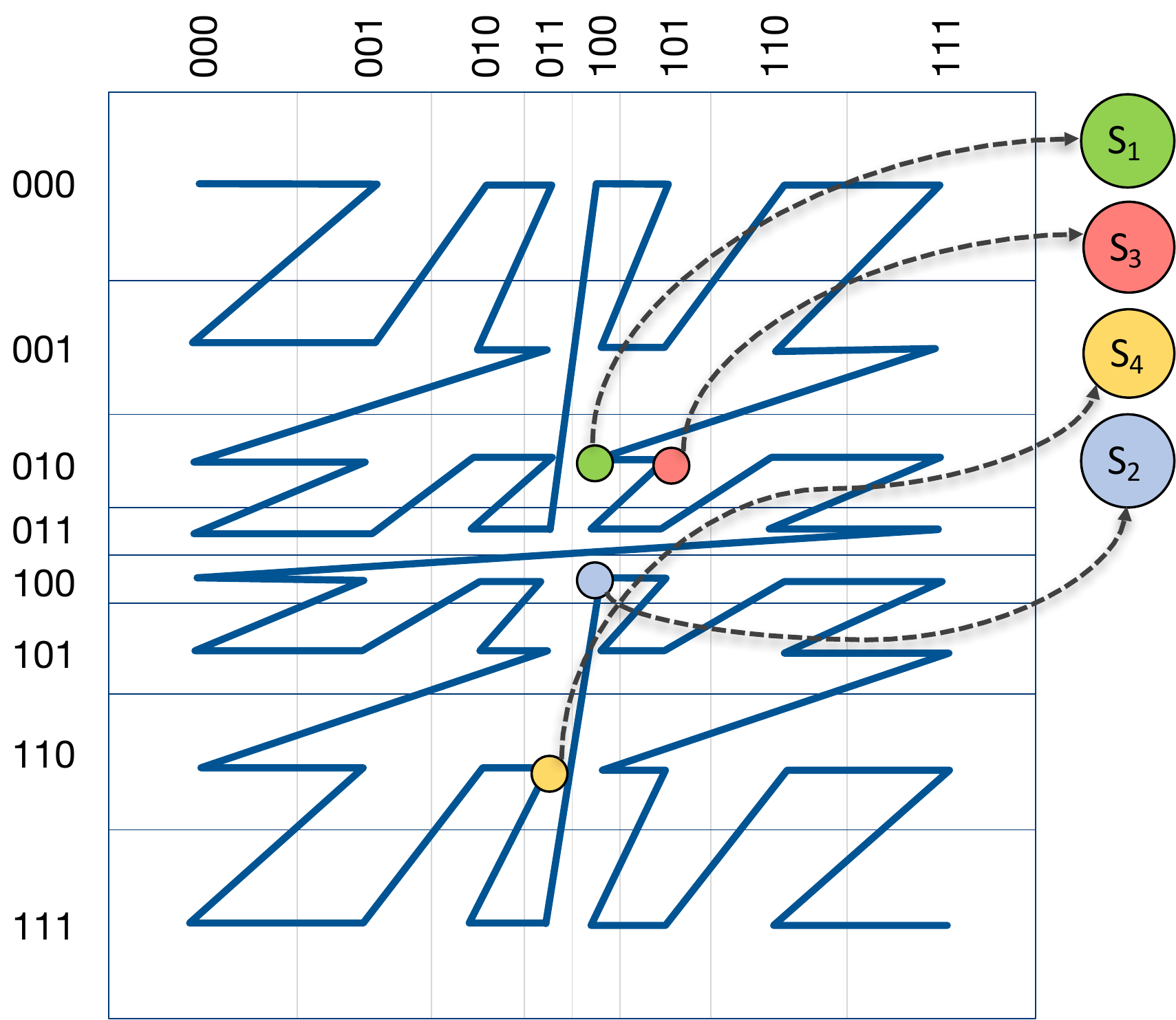}
	\caption{Z-ordered SAX summarization}
	\label{fig:sortingz}
\end{figure}

In Figure~\ref{fig:sortingz}, we show how to transform the four summarizations in Figure~\ref{fig:sorting} into sortable summarizations.
The key idea is placing the more significant bits first, while preserving them in the order of the segment that they belong to in the data series.
Note that this is a known technique used to project multi-dimensional data onto memory while preserving locality by storing them along a z-ordered space-filling curve~\cite{morton1966}.
An implementation of this technique for data series is shown in Algorithm~\ref{sortablesdrs}, transforming existing summarization schemes into sortable ones. To the best of our knowledge we are the first to apply this into data series summarizations.
In Figure~\ref{fig:sortingz}, we show the data series as points along the multi-dimensional space, where every segment represents a dimension. The figure also shows their linearized order along the z-ordered curve. As shown, the data series that are most similar to each other are indeed placed closest to each other (which is not the case when sorting them based on the original representation).

Note that a sortable summarization contains the same amount of information as the original summarization, the only difference being that the bits are ordered differently.
Hence, it is easy and efficient to switch back and forth between sortable summarizations and the original form, and we therefore do not lose anything in terms of the ability to prune the index during querying.

\Paragraph{New Infrastructure Opportunities} The ability to sort data series summarizations enables a plethora of new indexing infrastructure possibilities for data series indexes, ranging from read-optimized B-trees~\cite{Rao2000} to write-optimized LSM-trees~\cite{DBLP:journals/acta/ONeilCGO96} to adaptive structures that change performance characteristics based on workload~\cite{idreos2007, dayan2017}. Coconut-Trie and Coconut-Tree represent two points in this space that push upon the current state-of-the-art, though we expect that many more opportunities for specialization based on hardware and workload are possible.




\begin{algorithm}[tb]
\caption{Sortable Summarization}
\label{sortablesdrs}
\begin{algorithmic}[1]
	\small
\Procedure{invertSum}{Sum}
\For{each bit i of a segment in Sum}
    \For{each segment j}
         \State Add the i bit of segment j to SSum
    \EndFor
\EndFor
\State return SSum
\EndProcedure
\end{algorithmic}
\end{algorithm}


\subsection{Coconut-Trie}
\label{sec:trie}

We now present Coconut-Trie, a data series index that uses sortable summarizations to
construct a contiguous index using bulk-loading. Similarly to the state-of-the-art indexing schemes, Coconut-Trie divides data entries among nodes based on the greatest common prefix among all segments.
The advantage relative to the state-of-the-art is that
the resulting index is contiguous, meaning that queries do not issue random I/Os, but a large sequential I/O.

\Paragraph{Construction} The construction algorithm is shown in Algorithm~\ref{bottomup}.
The algorithm initially constructs the sortable summarizations of all data series and sorts them using external sort.
Then it constructs in a bottom-up fashion a detailed iSAX index.
Finally this index is compacted by pushing more data series in the leaf nodes.

\begin{algorithm}[tb]
\small

\caption{Coconut-Trie: bottom-up bulk-loading of an prefix split based tree}
\label{bottomup}
\begin{algorithmic}[1]
\Procedure{Coconut-Trie}{}
\While{not reached end of file}
   \State position = current file position;
   \State dataSeries= read data series of size n from file;
   \State SAX = convert dataSeries to SAX;
   \State invSAX = invertSum(SAX);
   \State Move file pointer n points;
   \State Add the (invSAX, position) pair in the FBL;
   \If{the main memory is full}
        \State Sort FBL according to invSAX
        \State Flush sorted FBL to the disk
   \EndIf
\EndWhile
\State Sort flushed runs using external sort

\While{not reached end of sorted file}
   \State Read the next (invSAX, position) in the FBL
   \If{the main memory is full}
        \For{every different subtree in FBL}
            \State //\textit{Move data from the FBL}
            \State //\textit{to leaf buffer}
            \State //\textit{and construct bottom-up the index}
            \For{every (invSAX, position) in FBL}
                 \State insertBottopUp(invSAX, position);
            \EndFor
            \State //\textit{merge leaf nodes as much as possible}
            \State CompactSubtree(root)
            \State //\textit{Flush all Leaf Buffers containing}
            \State //\textit{(Sax, position) pairs to the disk}
            \For{every leaf in subtree do}
                \State Flush the leaf to the disk;
            \EndFor
        \EndFor
  \EndIf
\EndWhile

\EndProcedure
\end{algorithmic}
\end{algorithm}

The input of the algorithm is a raw file, which contains all data series.
The process starts with a full scan of the raw data file in order to create the sortable summarizations for all data series (lines 4-6).
For data series we also record their offset in the raw file, so future queries can easily retrieve the raw values.
All sortable summarizations and offsets are stored in an FBL buffer (First Buffer Layer).
As soon as the buffer is full, it is sorted in the main memory and the sorted pairs are written to disk.
The process continues until we reach the end of the raw file.
If there are more than one sorted runs on disk, we sort them using external sort, and the final sorted file is written to disk.

Having the sortable summarizations sorted, all records that belong to a specific subtree are grouped together.
As such we exploit them in order to build a minimal tree in a bottom-up fashion, i.e., a tree that does not contain any raw data series (lines 22-24).
The main idea of the corresponding algorithm, i.e. the \textit{insertBottopUp} procedure, is that initially a new node is created for each different SAX representation.
Then, the algorithm replaces in iterations the least significant bits of the SAX representations with star marks until a common SAX prefix is identified to be placed in the parent node.
Then this idea is applied at the parent level and so on, until we reach the root (the corresponding algorithm is omitted due to lack of space).

The next phase is to compact this subtree, i.e. to push as many records in the leaf nodes as possible.
This is performed using the \textit{CompactSubtree} procedure (line 26).
To do that the algorithm iteratively checks whether the records of two sequential sibling nodes can fit together in a parent node.
If they do, the algorithm merges them and continues till all leaf nodes are visited.
Then the algorithm iterates again over the all leaves, until no more leaves are merged.
Finally each compacted subtree is flushed back to disk (lines 29-31).

The above algorithm is used to create a secondary index over the original raw file, keeping only the offsets in the leaf nodes. The algorithm performs the following passes over the data: (i) read the raw data series and compute the sortable summarizations; (ii) flush the sorted partitions of the summarizations to disk (along with their offsets); (iii) merge-sort them; and (iv) build the index.
This process involves $O(N/B)$ I/Os, but usually all the summarizations and their offsets fit in main memory, eliminating the need for passes (ii) and (iii).

A slight variation of the aforementioned algorithm could be used to create a fully-materialized iSAX index as well.\footnote{In a materialized index, the raw data-series are stored alongside their summarizations within the index, whereas in a non-materialized one the index contains \emph{pointers} to the raw data series that are stored in a different file.}
We call this variation \emph{Coconut-Trie-Full}.
This would require the raw data series to be sorted alongside their sortable summarizations in the sort-merge phase, and then flushed to disk.
Although the complexity of the algorithm would be the same, it would require additional passes in the sort-merge phase, and an additional pass over the raw data, in order to flush them to the leaf nodes.

\begin{figure}[tb]
    \includegraphics[width=\columnwidth]{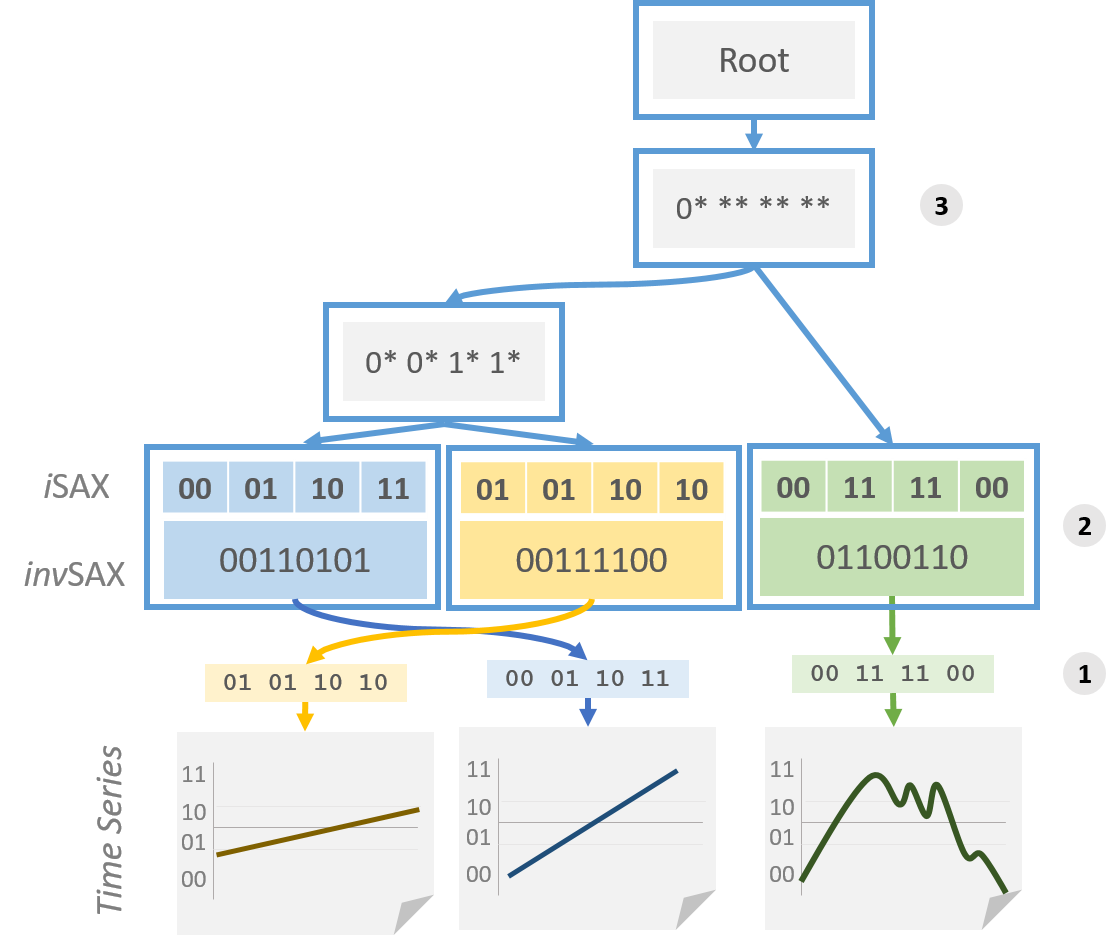}
    \caption{ Constructing bottom-up a Coconut-Trie index - before calling the \textit{compactSubtree} procedure. }
    \label{fig:example}
\end{figure}
\begin{exmp}
Figure~\ref{fig:example} illustrates an example of creating a Coconut-Trie index using the bottom-up Algorithm~\ref{bottomup}. As shown in the figure, we initially construct the summarizations (SAX) for all data series, as well as their sortable summarizations (invSAX).
Then, we sort them using their invSAX value, and we construct the corresponding Coconut-Trie index using the \textit{InsertBottomUp} algorithm.
Following this algorithm, initially, the first data series is placed in a new node.
The second data series is placed in yet a new node, since it has a different SAX representation than the first one.
Then, the \textit{createUptree} procedure is called to link the new node with the previous node. As such, the four least significant bits are replaced with stars, until the algorithm identifies a common prefix that could be used as the mask of the parent node $(0* 0* 1* 1*)$.
The parent is generated and linked to the root node.
The third data series is then inserted to the tree, and a new node is generated.
This node should be linked to the already existing tree: the \textit{createUptree} procedure is called again, using as input the SAX representations of the second and third data series.
The least significant bits are again replaced by a star, one by one until we identify the parent that should be generated linking the third node to the tree.
The resulting Coconut-Trie tree (refer to Figure~\ref{fig:example}) demonstrates the state of the tree before calling the \textit{CompactSubtree} procedure, which will follow in order to compact the entire tree.
Assuming that a leaf node can hold two data series, the corresponding algorithm will identify that the first two time-series have the same parent and they fit together.
As such they can be placed directly in their parent node, removing the child nodes.
\end{exmp}

\Paragraph{Queries} Since the constructed index is essentially no different than an iSAX index, we use the traditional approximate and exact search algorithms in order to perform querying.
Approximate search works by visiting the single most promising leaf, and calculating the minimum distance to the raw data series contained in it.
It provides answers of good quality (returns a top 100 answer for the nearest neighbor search in 91.5\% of the cases for iSAX with extremely fast response times~\cite{Shieh2008, Shieh2009}).
On the other hand, exact search guarantees that we get the exact answer, but with potentially much higher execution time.
For exact search, we employee the SIMS algorithm, implementing a skip sequential scan algorithm, shown to outperform traditional exact search algorithms~\cite{ZoumpatianosIP16}.

\subsection{Coconut-Tree}
\label{sec:tree}

Although Coconut-Trie achieves contiguity, i.e. the records placed in each leaf are close in the hyperspace, a lot of disk space is wasted in those leafs: many of them are only half-full, due to the way the index is constructed (i.e., compacting child nodes to a parent one).
In addition, since the constructed tree in both Coconut-Trie and in current state-of-the-art are unbalanced trees, they offer no guarantees for the query answering time.

We now present Coconut-Tree, a data series index that organizes data series based on sortable summarizations, and improves upon Coconut-Trie by eliminating the constraint that a node can only contain elements with a common prefix.
This leads to a balanced index that can densely pack data in its leaf nodes (at a fill-factor that can be controlled by the user).
The corresponding algorithm completes index construction again in $O(N/B)$ time.

The algorithm is shown in Algorithm~\ref{ubtree} and gets as input again the raw data file.
A buffer is initialized, and while the buffer is not full the next data series is loaded from the raw file, and the sortable summarization is calculated and stored along with the position of this data series in the raw data file (lines 2-8).
Then this buffer is sorted using external sort (line 9-14), and the UB-Tree bulk-loading algorithm~\cite{DBLP:books/daglib/0011128} is called to construct the final index.
This algorithm requires a sorted input, and starts by filling in the lead records and constructing the parent nodes by using median-based splits.

\begin{algorithm}[tb]
	\caption{Coconut-Tree: Bottom-up bulk-loading of a balanced tree}
	\label{ubtree}
	\begin{algorithmic}[1]
		\small
		\Procedure{Coconut-Tree}{}
		\While{not reached end of file}
		\State position = current file position;
		\State dataSeries = read data series of size n from file;
		\State iSAX = convert dataSeries to iSAX;
		\State invSAX = invertSum(iSAX);
		\State Move file pointer n points;
		\State Add the (invSAX, position) pair in the FBL;
		\If{the main memory is full}
		\State Sort FBL according to invSAX
		\State Flush sorted FBL to the disk
		\EndIf
		\EndWhile
		\State Sort flushed runs using external sort
		\State Use UB-Tree bulk-loading algorithm to build a tree on top of the sorted file.
		\EndProcedure
	\end{algorithmic}
\end{algorithm}

The Algorithm~\ref{ubtree} builds a secondary index with only offsets in the lead nodes, but it can be used to construct a fully materialized index as well, where all data reside in the leaf nodes.
We call the materialized version of the algorithm \emph{Coconut-Tree-Full}.
We expect that index construction time of Coconut-Tree-Full will be significantly larger.
Nevertheless, we also expect that query execution time would be better, since it will not perform additional I/Os to go to the raw data file for accessing each required data series record.

\begin{figure}[tb]
	\centering
    \includegraphics[width=\columnwidth]{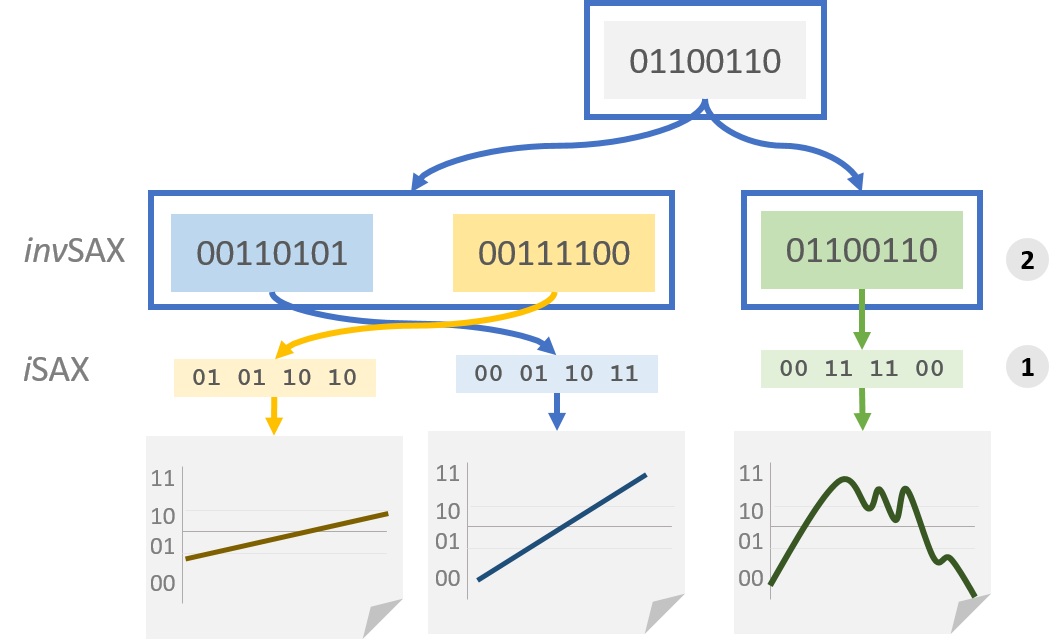}
    \caption{ Constructing a Coconut-Tree index. }
    \label{fig:examplecoconut}
\end{figure}

\begin{exmp}
Figure~\ref{fig:examplecoconut} illustrates the construction of a Coconut-Tree index.
Initially, we construct for all data series their SAX and their invSAX representations.
We then sort them using their invSAX value, and we construct the Coconut-Tree index in a bottom-up fashion (exploiting the bulk-loading algorithm for UB-Trees~\cite{DBLP:books/daglib/0011128}).
Note that the constructed index in this case is balanced.
\end{exmp}

\Paragraph{Querying} For \textit{approximate search}, when a query arrives (in the form of a data series), it is first converted to its sortable summarization.
Then the Coconut-Tree index is traversed searching for this sortable summarization similar the approximate search in iSAX trees.
The idea is to search for the leaf, where the query series would reside if it was part of the indexed data set.
If such a record exists, it is retrieved from the disk and returned to the user.
On the other hand, if such a record does not exist, all data series in a specific radius from this specific point are retrieved from the disk (usually a disk page), and their real distances from the query are calculated.
The data series with the minimum distance found among the data series in that radius is used as the approximate answer.
Thus, in terms of execution cost, the algorithm visits as many nodes as the depth of the tree, and any additional leaf nodes within the selected radius.

Note that in a Coconut-Tree index, we have pointers between neighboring leaves, which are allocated sequentially on disk.
This allowed us to experiment with the radius size, optimizing the trade-off between the quality of the answer and the execution time of the approximate search.

 \begin{algorithm}[tb]
    \caption{Approximate search for the Coconut-Tree}
    \label{approxsearch}
    \begin{algorithmic}[1]
    	\small
\Procedure{approxSearchCoconutTree}{dataSeries, invSAX, index, radius}
        \State targetPoint = point where invSAX should be inserted
        \State //\textit{Calculate the real leaf distance between }
        \State //\textit{the dataSeries and the raw data series }
        \State //\textit{in a radius around the place that the }
        \State //\textit{dataSeries should reside if existed}
        \State bsf = caclRadLeafDist(targetPoint, dataSeries, radius);
\EndProcedure
    \end{algorithmic}
\end{algorithm}

For implementing \textit{exact search} for Coconut-Tree, we implement a skip sequential scan algorithm (refer to Algorithm~\ref{exactsearch}) similar to SIMS~\cite{ZoumpatianosIP16}.
Our algorithm employs approximate search as a first step in order to prune the search space.
It then accesses the data in a sequential manner, and finally produces an exact, correct answer. We call this algorithm Coconut-Tree Scan of In-Memory Summarizations (CoconutTreeSIMS).
The main intuition is that while the raw data do not fit in main memory, their summarized representations (which are orders of magnitude smaller) will fit in main memory (remember that the SAX summaries of 1 billion data series occupy merely 16 GB in main memory). By keeping these data in-memory and scanning them, we can estimate a bound for every data series in the data set.

 \begin{algorithm}[tb]
    \caption{Coconut-Tree Scan of In-Memory summarizations }
    \label{exactsearch}
    \begin{algorithmic}[1]
    	\small
\Procedure{coconutTreeSIMS}{dataSeries, invSAX, index, radius}
        \State //\textit{if SAX sums are not in memory, load them}
        \If{invSums = 0}
          \State invSums = loadinvSaxFromDisk();
        \EndIf
        \State //\textit{perform an approximate search}
        \State bsf = approxSearchCoconutTree(dataSeries, invSAX, index, radius);
        \State //\textit{Compute minimum distances for all summaries}
        \State Initialize mindists[] array;
        \State //\textit{use multiple threads \& compute bounds in parallel}
        \State parallelMinDists(mindists, invSums, dataSeries);
        \State //\textit{Read raw data for unprunable recorde}
        \State recordPosition = 0;
         \For{every mindist in mindists}
            \If{mindist < bsf}
                  \State rawData = read raw data series from index;
                  \State realDist = Dist(rawData, dataSeries);
                  \If{realDist < bsf}
                    \State bsf = realDist;
                  \EndIf
            \EndIf
            \State recordPosition++;
        \EndFor

\EndProcedure
    \end{algorithmic}
\end{algorithm}

The algorithm differs from the original SIMS algorithm in that it searches over the sorted invSAX representations for the initial pruning, and it then uses the Coconut-Tree index to get the raw data-series instead of accessing the original file with the raw data series.
As such, Algorithm~\ref{exactsearch} starts by checking whether the sortable summarization data are in memory (lines 3-4), and if not it loads them in order to avoid recalculating them for each query.
It then creates an initial best-so-far (bsf) answer (line 7), using the approximate search algorithm described previously (Algorithm~\ref{approxsearch}).
A minimum distance estimation is calculated between the query and each in-memory sortable summarization (line 11) using multiple parallel threads, operating on different data subsets.
For each lower bound distance estimation, if this is smaller than the real distance to the bsf, we fetch the complete data series from the Coconut-Tree index, and calculate the real distance (lines 15-22).
If the real distance is smaller than the bsf, we update the bsf value (lines 19-21).
Since the summaries array is aligned to the data on disk, what we essentially do is a synchronized skip sequential scan of the raw data and the in-memory mindists array.
This property allows us to prune a large amount of data, while ensuring that the executed operations are very efficient: we do sequential reads in both main memory and on disk, and we use modern multi-core CPUs to operate in parallel on the data stored in main memory.
At the end, the algorithm returns the final bsf to the user, which is the exact query answer.


\section{Experimental Evaluation}

In this section, we present our experimental evaluation.
We demonstrate the benefits of the sortability, enabling a variety of choices for data structures to be used for efficiently bulk-loading data series.
To this end, we show that Coconut-Tree, overall, has better resilience when the data volume increases significantly with respect to the main memory size.
In addition, we show that Coconut-Tree is more efficient in query answering.

\Paragraph{Algorithms}
We benchmark all indexing methods presented in this paper, and we compare index building and query answering performance, to the current state-of-the-art.
More specifically, we compare our materialized methods with R-tree~\cite{Guttman1984},  Vertical~\cite{DBLP:conf/kdd/KashyapK11}, DSTree~\cite{Wang2013} and ADS-Full~\cite{ZoumpatianosIP16}, and and our non-materialized methods with ADS+~\cite{ZoumpatianosIP16} and a non-materialized version we implemented over R-tree, the R-tree+.

The Vertical approach generates an index using data series features, obtained by a multi-resolution Discrete Wavelet Transform, in a stepwise sequential-scan manner, one level of resolution at a time.
DSTree is a data adaptive and dynamic segmentation tree index that provides tight upper and lower bounds on distances between time series.
ADS-Full is an algorithm that constructs an $i$SAX style clustered index by performing two passes over the raw data series file.
ADS+ is an adaptive data structure, which starts by building a minimal secondary index.
Leaf sizes are refined during query answering, and leaves are materialized on-the-fly.
We consider ADS+ a non-materialized index.
The R-tree index is built on the raw data series by indexing their PAA summarizations.
The raw data series are stored in the leaves of the tree.
Our R-tree implementation uses the Sort-Tail-Recursive bulk loading algorithm~\cite{leutenegger1997}.
R-tree+ is the non-materialized version of the R-tree, using file pointers in the leaves instead of the original time series.
In our experiments, we used the same leaf size (2000 records) for all indexing structures.

\Paragraph{Infrastructure}
All algorithms are compiled with GCC 4.6.3 under Ubuntu Linux 12.04 LTS. We used an Intel Xeon machine with
5x2TB SATA 7.2 RPM hard drives in RAID 0.
The memory made available for each algorithm was varied according to the experiment.

\Paragraph{Datasets}
\begin{figure}[tb]
	\begin{center}
		\includegraphics[width=\columnwidth]{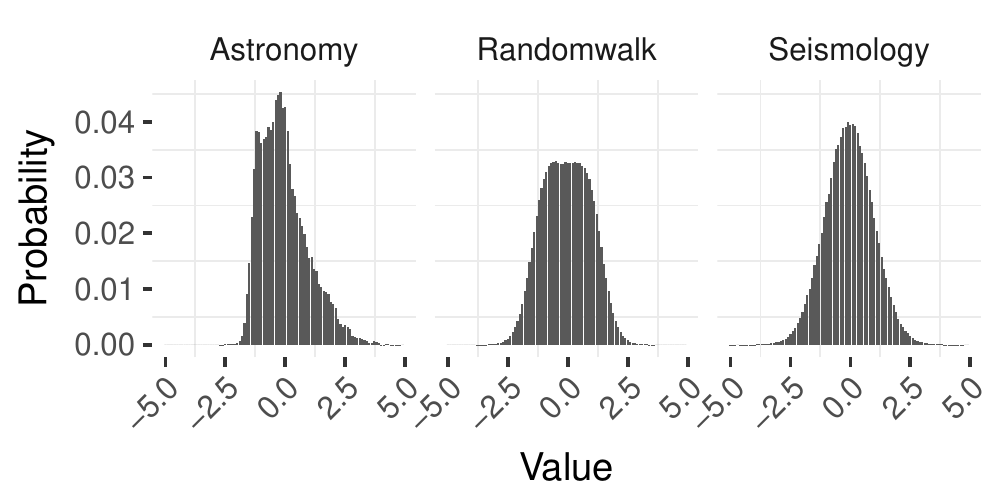}
		\caption{Value histograms for all datasets used}
		\label{fig:distributions}
	\end{center}
\end{figure}
For our experiments we used both synthetic and real datasets. Synthetic datasets were generated using a random walk data series generator: a random number is drawn from a Gaussian distribution (0,1); then, at each time point a new number is drawn from this distribution and added to the value of the last number.
This kind of data has been extensively used in the past (see~\cite{ZoumpatianosLPG15} for a list of references), and has been shown to effectively model real-world financial data~\cite{DBLP:conf/sigmod/FaloutsosRM94}.

The real datasets we used in our experiments are seismic and astronomy data.
We used the IRIS Seismic Data Access repository~\cite{iris} to gather data series representing seismic waves from various locations.
We obtained 100 million data series of size 256 using a sliding window with a resolution of 1 sample per second, sliding every 4 seconds. The complete dataset size was 100GB.
For the second real dataset, we used astronomy data series representing celestial objects~\cite{refId0}.
The dataset comprised of 270 million data series of size 256, obtained using a sliding window with a step of 1. The total dataset size was 277GB.

All our datasets have been z-normalized by subtracting the mean and dividing by the standard deviation.
This is a requirement by many applications that need to measure similarity irrespective of translation and scaling of the data series.
Moreover, it allows us to compute correlations based on the ED values~\cite{MueenNL10}.
In Figure~\ref{fig:distributions}, we show the distributions of the values for all datasets.
The distributions of the synthetic and seismology data are very similar, while astronomy data are slightly skewed.

\Paragraph{Workloads}
The query workloads for every scenario are random. Each query is given in the form of a data series $q$ and the index is trying to locate whether this data series or a similar one exists in the database.
For querying the real datasets we obtained additional data series from the raw datasets using the same technique for collecting the datasets to be used in the query workload.

\subsection{Indexing}

\begin{figure*}[tb]
	\begin{subfigure}{0.32\textwidth}
		\begin{center}
			\includegraphics[width=\textwidth]{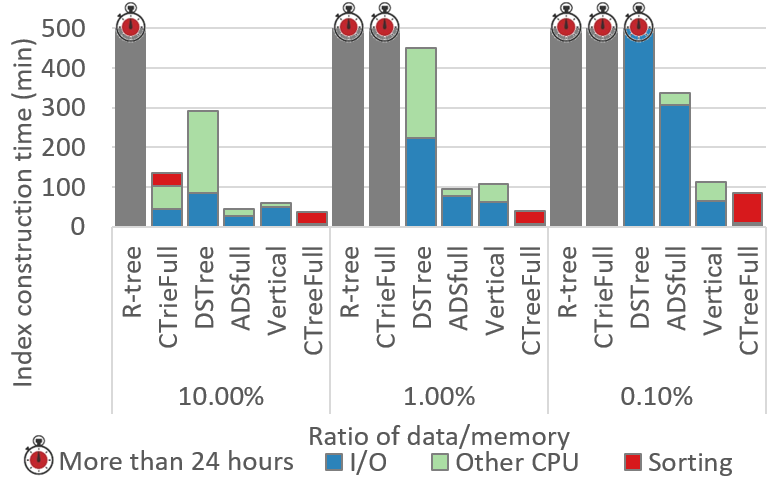}
			\caption{Index construction - materialized}
			\label{fig:fixdata}
		\end{center}
	\end{subfigure}
	\begin{subfigure}{0.32\textwidth}
		\begin{center}
			\includegraphics[width=\textwidth]{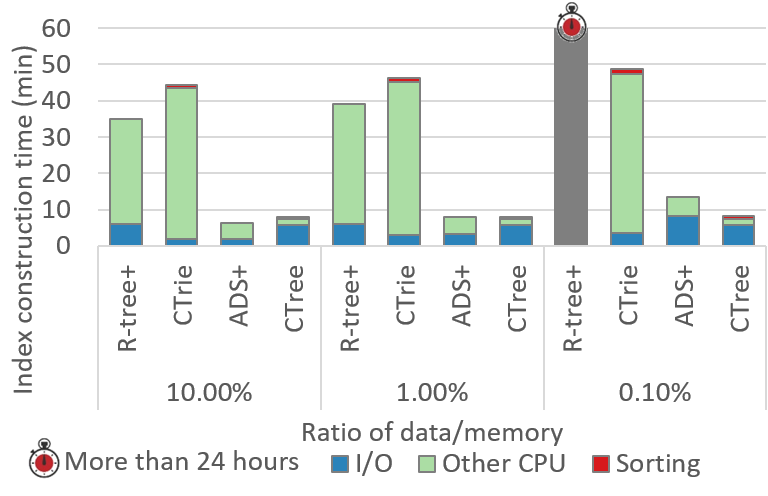}
			\caption{Index construction - non-materialized }
			\label{fig:fixdata+}
		\end{center}
	\end{subfigure}
\begin{subfigure}{0.32\textwidth}
	\begin{center}
		\includegraphics[width=\textwidth]{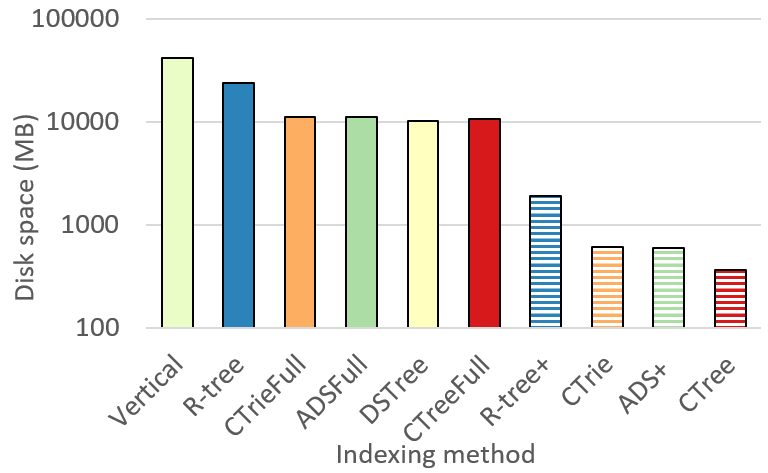}
		\caption{Indexing space overhead.}
		\label{fig:space}
	\end{center}
\end{subfigure}
\begin{subfigure}{0.32\textwidth}
	\begin{center}
		\includegraphics[width=\textwidth]{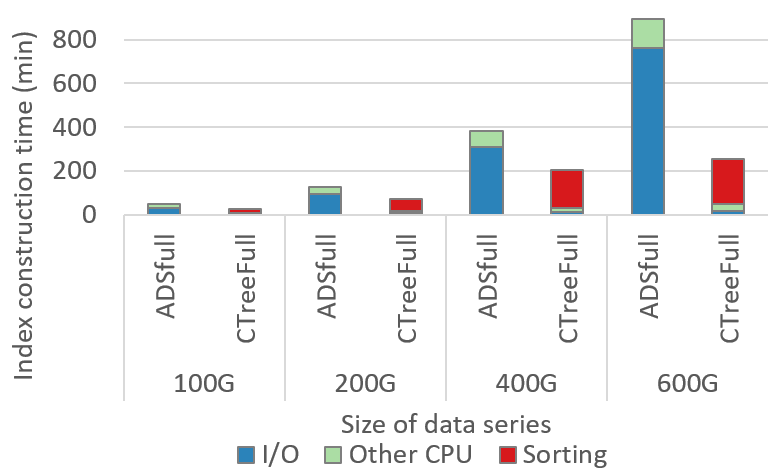}
		\caption{Index construction - materialized. }
		\label{fig:fixmemory}
	\end{center}
\end{subfigure}
\begin{subfigure}{0.32\textwidth}
	\begin{center}
		\includegraphics[width=\textwidth]{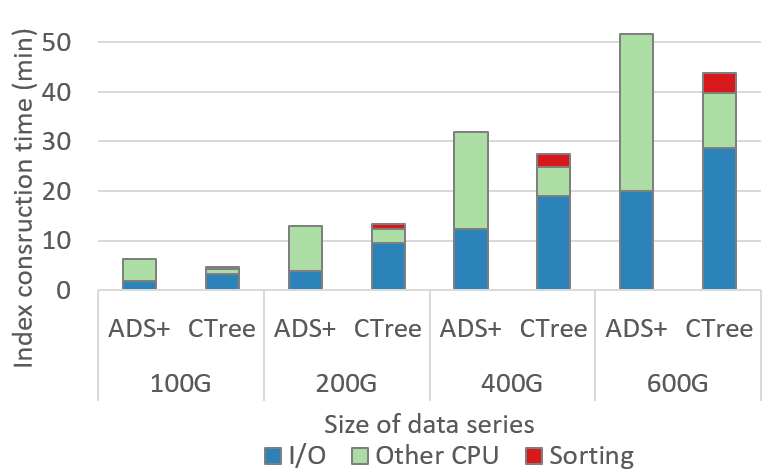}
		\caption{Index construction - non-materialized.}
		\label{fig:fixmemory+}
	\end{center}
\end{subfigure}
	\begin{subfigure}{0.32\textwidth}
	\begin{center}
		\includegraphics[width=\textwidth]{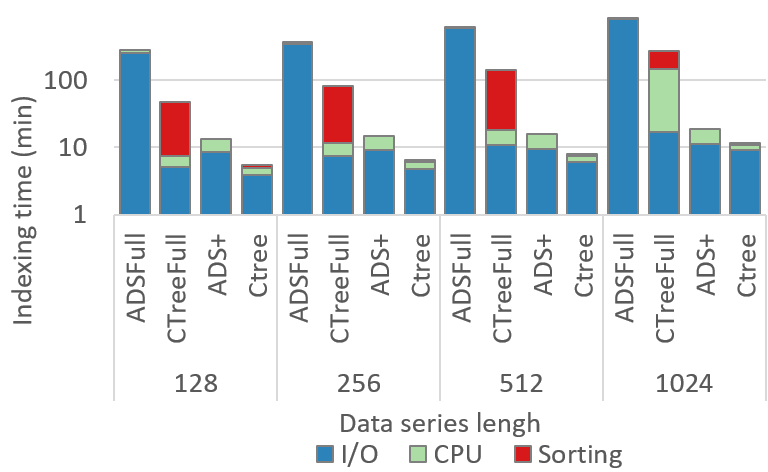}
		\caption{Indexing variable length data series.}
		\label{fig:variable}
	\end{center}
\end{subfigure}
	\caption{Indexing}
\end{figure*}



In our first set of experiments, we evaluate index construction. The results for the materialized algorithms are shown in Figure~\ref{fig:fixdata}. As we can observe, even with ample memory, Coconut-Tree-Full (CTreeFull) exhibits the best performance.
In addition, most of the construction time is spent on external sorting of the raw data file.
As expected, when the memory becomes limited, external sorting requires more time to sort the raw data files.
The execution time of Coconut-Trie-Full (CTrieFull) on the other hand, significantly increases as we constrain the memory (and the corresponding buffering), due to the extensive I/Os spent on the last pass of data, for loading the unsorted raw data to the sorted leaves.
Moreover, we observe that Vertical is slower in all cases, 
while R-tree performs rather poorly.
The STR algorithm~\cite{leutenegger1997} that R-tree uses first sorts based on the first dimension into $N^{\frac{1}{D}}$ slabs (where $N$ is the number of points in a $D$-dimensional space), and then recursively repeats the process within each slab with one less dimension. 
As a result, runtime is the product of the number of elements and the number of dimensions: $O(N \cdot D)$ I/Os. 
In contrast, our implementation uses sortable summarizations to sort based on all dimensions with just one pass, amounting to $O(N)$ I/Os.
Finally, DSTree requires more than 24 hours to finish in most of the cases, as it inserts all data series in the index one by one, in a top-down fashion.
This requires multiple iterations to be performed over the raw data during splits in order to create more detailed summarizations, leading to a high I/O overhead.

In the non-materialized versions of the algorithms, shown in Figure~\ref{fig:fixdata+}, ADS+ is slightly better than Coconut-Tree (6.3 vs 7.8 mins), when given enough memory.
However when we restrict the available main memory, Coconut-Tree eventually becomes faster than ADS+ (8.2 vs 13.4 mins).
This is due to the fact that as the leaves in ADS+ split, they cause many small random disk I/Os.
This significantly slows down index construction, since buffering is limited when the main memory is limited.
On the other hand, Coconut-Trie (CTrie) spends a significant time in compacting its nodes, which significantly slows down index construction.
The performance of R-tree+ matches the behavior of the materialized R-tree, requiring much more execution time than the leading approaches.

Finally, we observe that non-materialized versions outperform the materialized ones, since they do not store the entire dataset, but only the summarizations and pointers to the raw data file.
Moreover, we note that sorting in the non-materialized versions is really fast, since only the summarizations need to be sorted, which in general fit in main memory.

\Paragraph{Space} Since space overhead is critical in many scenarios, next we examine the space overhead imposed by the various indexing schemes. The results are shown in Figure~\ref{fig:space}, where we report the space required for answering queries over 10GB of raw data.

For the materialized indexes, we observe that Coconut-Tree-Full and DSTree impose the smaller space overhead. Median-based solutions, such as Coconut-Tree-Full generate indexes with the leaf nodes as full as possible, whereas in prefix-based solutions there is a lot of empty space in the leaf nodes: leaves are on average 10\% full in prefix-based solutions, whereas for the median-based ones utilization reaches 97\%.
Note that in the case of Coconut-Trie-Full more space is wasted, since more leaf nodes are produced, and we cannot compact any more the leaf nodes due to the specific prefix-based scheme that is used (there are 55K leaf nodes for the Coconut-Trie-Full, and 54K leaf nodes for the ADSFull).

For the non-materialized indexes, we can again observe the superiority of our median based solution, requiring almost half the space required by other solutions.

\Paragraph{Fixed amount of main memory}
Having selected CTreeFull as out proposed solution, 
we proceed in the evaluation only with the Coconut-Tree and the ADS families of algorithms.
In this set of experiments, we fix the amount of main memory to that of a common desktop workstation (8GB), and gradually increase the number of data series to be indexed. The results are shown in Figures~\ref{fig:fixmemory} and~\ref{fig:fixmemory+}.
We observe that when the amount of data is relatively small with respect to the available main memory, Coconut-Tree-Full and Coconut-Tree require similar times to ADSFull and ADS+, respectively.
However, as the data size gradually increases, the random I/Os of ADSFull and ADS+ incur a significant overhead on the overall time to construct the index, and the Coconut-Tree algorithms become faster. In addition, the experiments show that in Coconut-Tree-Full most of the time is spent on sorting the raw data, whereas in the case of Coconut-Tree only the summarizations are sorted, and as such the external sort overhead is really small when compared to the cost of I/Os and CPU.
\begin{figure*}[tb]
	\begin{subfigure}{0.32\textwidth}
		\begin{center}
			\includegraphics[width=\textwidth]{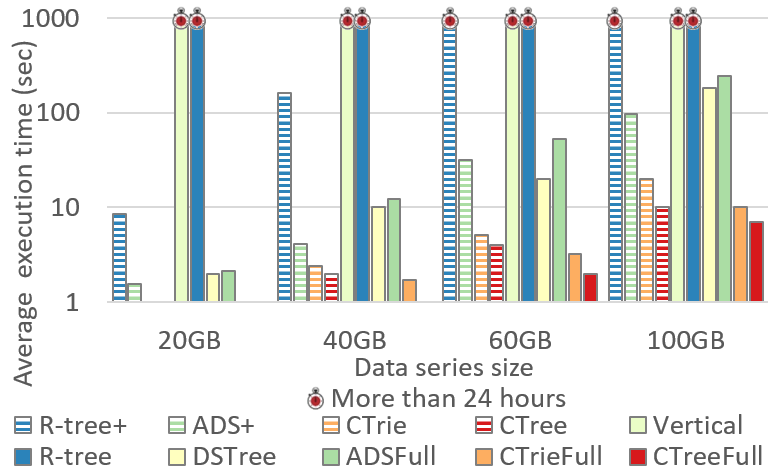}
			\caption{Exact query answering }
			\label{fig:exact}
		\end{center}
	\end{subfigure}
	\begin{subfigure}{0.32\textwidth}
		\begin{center}
			\includegraphics[width=\textwidth]{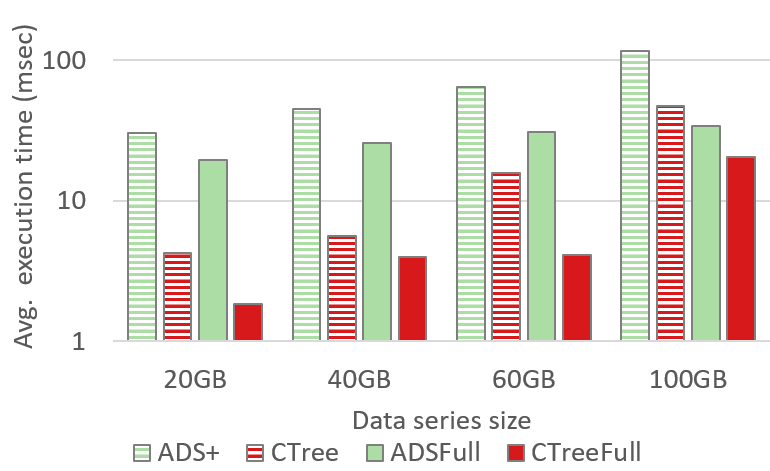}
			\caption{Approximate query answering.}
			\label{fig:approximate}
		\end{center}
	\end{subfigure}
	\begin{subfigure}{0.32\textwidth}
		\includegraphics[width=\textwidth]{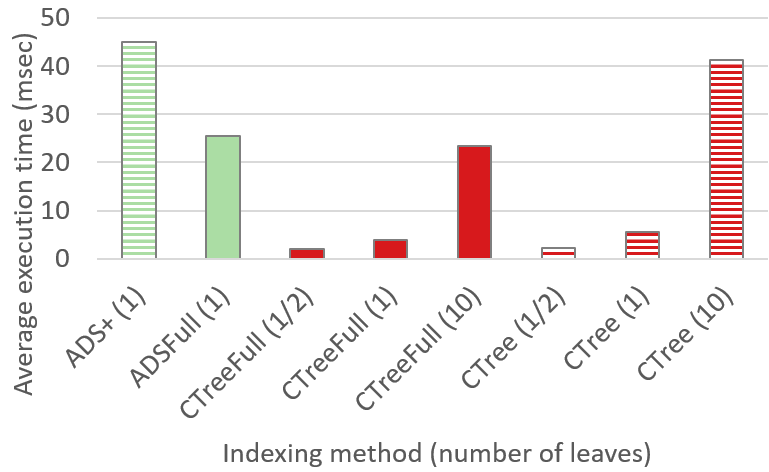}
		\caption{Approximate query answering (40G).}
		\label{fig:quality_approximate}
	\end{subfigure}
	\begin{subfigure}{0.32\textwidth}
		\includegraphics[width=\textwidth]{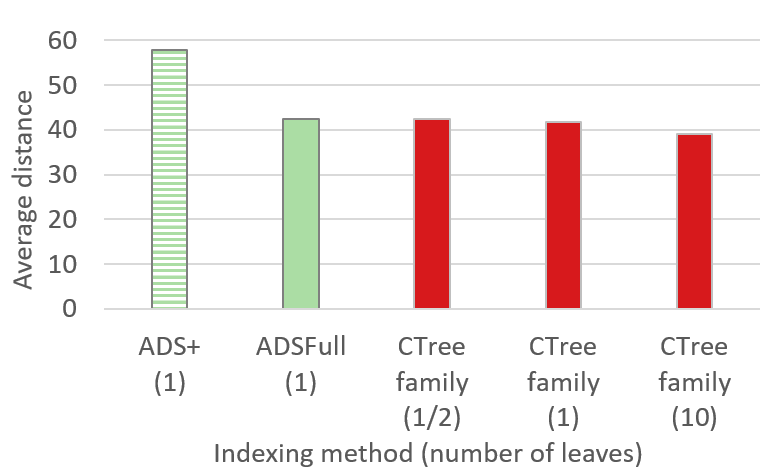}
		\caption{Average distance of approximate search.}
		\label{fig:quality_avdist}
	\end{subfigure}
	\begin{subfigure}{0.32\textwidth}
		\includegraphics[width=\textwidth]{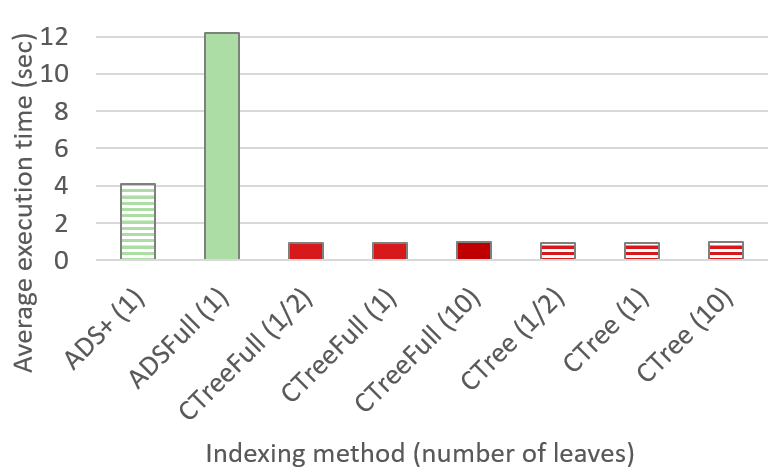}
		\caption{Exact query answering.}
		\label{fig:quality_exact}
	\end{subfigure}
	\begin{subfigure}{0.32\textwidth}
		\includegraphics[width=\textwidth]{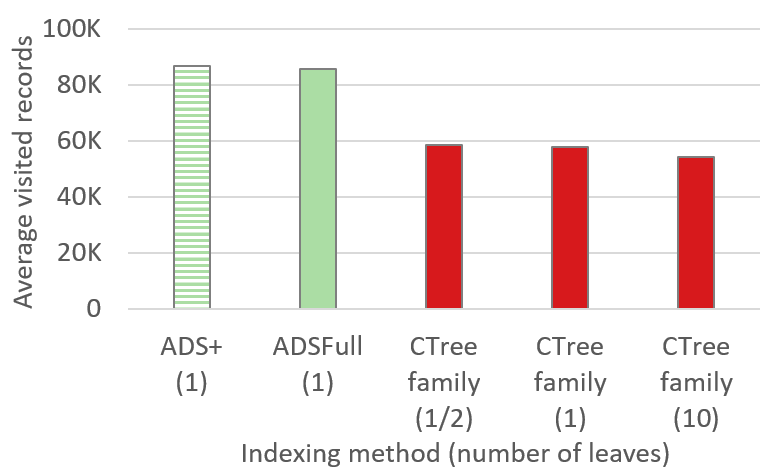}
		\caption{Visited records in exact query answering.}
		\label{fig:quality_visited records}
	\end{subfigure}
	\caption{Querying.}
\end{figure*}

\Paragraph{Variable length of data series}

Finally, we study the behavior of indexing data series collections of 100GB, using limited memory (100K data series) and different series lengths.
The results are shown in Figure~\ref{fig:variable}.
We observe that in all cases the Coconut-Tree variations surpass the ADS ones, demonstrating once again the superiority of Coconut-Tree.

\subsection{Querying}
\Paragraph{Exact Query Answering} Next we evaluate the performance of the various indexing schemes in terms of exact query answering.
To do that, we use 100 random queries over the constructed indexes for variable data size. As shown in Figure~\ref{fig:exact}, Coconut-Tree-Full and Cococut-Tree are faster for exact search.
This is because the corresponding indexes are contiguous and compact.
We observe that Coconut-Tree and Coconut-Tree-Full outperform in all cases the other solutions.
This is reasonable, as exact querying in all cases proceeds by first executing an approximate search.

Since the result of the approximate search has better quality (i.e., smaller Euclidean distance), less records need to be visited in the subsequent steps of exact search. We elaborate on this observation later in this section. An interesting observation here is that the non-materialized version of the R-tree in 40GB is faster than the materialized one. This happens since R-tree+ needs only the summarizations in memory to perform query answering, whereas the materialized version needs large parts of data series, which leads to memory swapping to disk.

\Paragraph{Approximate Query Answering}
We now evaluate the performance of the various indexes in terms of approximate query answering.
To do that, we use 100 random queries over the constructed indexes for variable data size. Again we focus only on the most promising indexing schemes for the rest of our experiments on querying.
The results are shown in Figure~\ref{fig:approximate}.
We can observe, that Coconut-Tree and Coconut-Tree-Full are always faster than the other methods.
In addition, the materialized versions of the indexes are faster in approximate query answering, since the records are materialized in the leaf nodes and can be directly accessed instead of going to the raw data file.


\Paragraph{Quality}
In order to further investigate exact query answering among the various indexing methods, we used 40GB of data series.
Since the first step of the exact search is the execution of an approximate query, a better initial approximate result leads to increased pruning and to visiting a smaller number of nodes.
Indeed, besides the fact that approximate query answering is faster in the Coconut-Tree family (refer to Figure~\ref{fig:quality_approximate}), the results of the approximate search have better quality, as well. This is shown in Figure~\ref{fig:quality_avdist}, where we observe that the average Euclidean distance between the input queries and the results of the approximate search, is smaller for the Coconut family algorithms.
In addition, CTree(1) produced better results than ADSFull for 69\% of the queries, and CTree(10) for 94\% of them.

Evidently, this has a direct impact on exact search, as illustrated in Figure~\ref{fig:quality_exact}.
Remember that both SIMS and CoconutTreeSIMS initially perform an approximate search over the dataset, and use this result to prune the search space.
Consequently, the better the result of approximate search, the more records are pruned in the subsequent phase of exact query answering.
Figure~\ref{fig:quality_visited records} shows that (on average) the ADS family visits more than 80K records during exact query answering, whereas the Coconut family visits less than 59K records in all cases.
By being able to visit multiple leaf nodes in the first (approximate search) step of CoconutTreeSIMS, we can identify a better initial answer, and further reduce the average number of visited records
(refer to Figure~\ref{fig:quality_avdist}).
However, this has an unexpected impact on exact search time performance: although more records are pruned (refer to Figure~\ref{fig:quality_visited records}), there is no benefit due the time spent for visiting more leaf nodes in the approximate search, as shown in Figure~\ref{fig:quality_exact}.

\subsection{Updates \& Complete Workloads}

\begin{figure*}[tb]
	\begin{subfigure}{0.32\textwidth}
		\includegraphics[width=\textwidth]{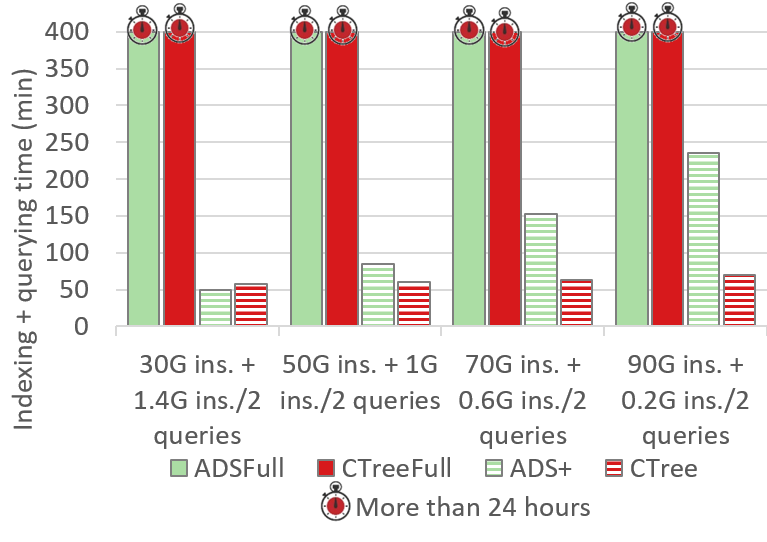}
		\caption{Mixed workload.}
		\label{fig:updates}
	\end{subfigure}
	\begin{subfigure}{0.32\textwidth}
		\includegraphics[width=\textwidth]{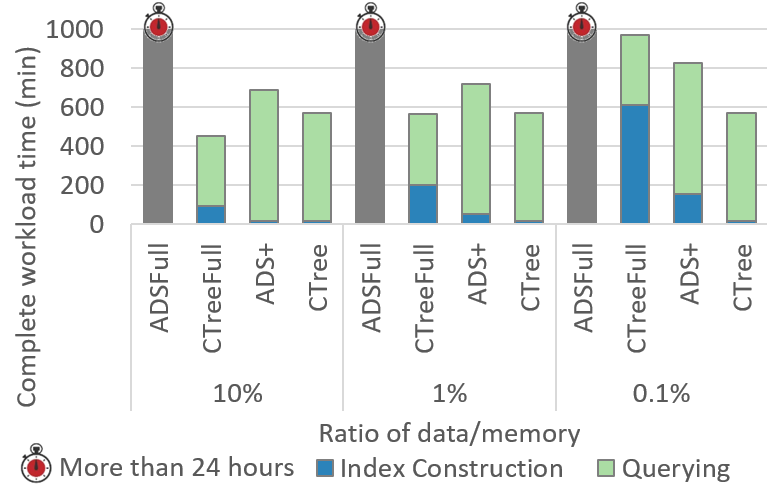}
		\caption{Astronomy - complete workload.}
		\label{fig:real_astro}
	\end{subfigure}
	\begin{subfigure}{0.32\textwidth}
		\includegraphics[width=\textwidth]{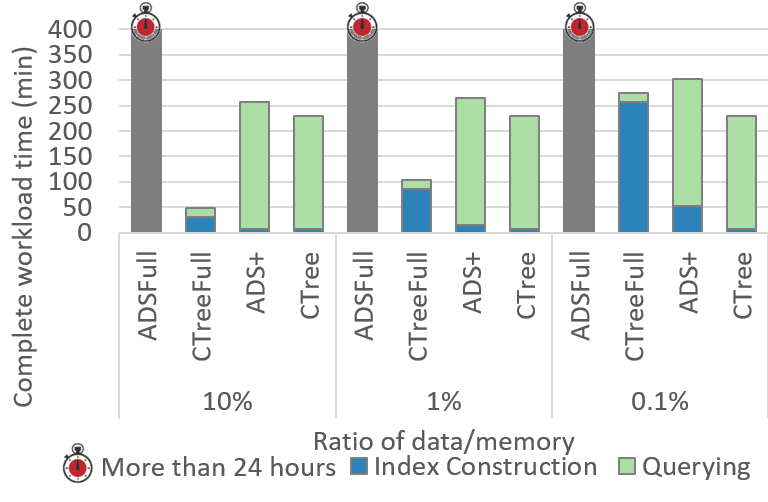}
		\caption{Seismic - complete workload.}
		\label{fig:real_seismic}
	\end{subfigure}
	\caption{Updates \& Complete Workloads.}
\end{figure*}

\Paragraph{Updates}
In our next experiment, we study the behavior of the ADS and the Coconut-Tree families under updates. We use a synthetic data set of 100GB of data series and 100 random exact queries. In addition, we limit the available memory (0.01\% of the available data). This time, queries are interleaved with updates. After an initial bulk loading of a varying number of data series, a batch of new data series arrives, followed by 2 queries. The batch and the queries are sequentially executed until we reach in total 100GB data series loaded in the index, and 100 exact queries executed in total. The results in Figure~\ref{fig:updates} show that when updates are highly fragmented, the ADS family behaves better. However, as we bulk load larger volumes of data series, CTree is the winner, because our bulk loading algorithm has to perform less splits when larger pieces of data are loaded.

\Paragraph{Real Datasets}
Finally, we compare Coconut to the state-of-the-art, simulating the complete process of index construction and query answering.
The results are shown in Figure~\ref{fig:real_astro} for the Astronomy dataset, and in Figure~\ref{fig:real_seismic} for the Seismic dataset.

The index sizes for the astronomy dataset were as follows: ADSFull: 311GB, ADS+: 19GB, CTree: 10GB, CTreeFull: 298GB; and for the seismic dataset: ADSFull: 111GB, ADS+: 6GB, CTree: 4GB, CTreeFull: 108GB.

We measure the time to construct first the corresponding indexes, and then to answer 100 exact queries over the constructed index, using various memory configurations.
As shown, when we constrain the available memory, Coconut-Tree becomes better in all cases, for both the materialized and non-materialized approaches, corroborating the experimental results with the synthetic datasets.
An interesting observation here is that the queries are harder on these datasets for all indexes, because the datasets were denser (for a detailed discussion of hardness see~\cite{ZoumpatianosLPG15}). 
As a result, pruning was not as efficient as with the random walk data. 
Therefore, even though Coconut was faster than all competing methods, it still had to scan a considerable amount of data in order to answer the exact queries.


\section{Conclusions and Future Work}

In this paper, we show that state-of-the-art data series indexing approaches cannot really scale, as the data size becomes significant larger than the available memory, exhibiting significant performance overheads.
We identify as a key problem the fact that existing summarizations cannot be effectively sorted.
To alleviate this problem, we propose the first \emph{sortable} data series summarizations, showing that indexing based on sortable summarizations optimizes both indexing and querying.
We start by creating and exploring a prefix-based bottom-up indexing algorithm,
which merely solve the problem of data contiguity.
We proceed 
by exploring median-based split trees, and showing that this approach outperforms the state-of-the-art for both index construction and querying time.
Among the benefits of the approach is that the resulting index structure is balanced, providing guarantees on query execution time.
As future work, 
we would also like to explore how ideas from LSM trees~\cite{DBLP:journals/acta/ONeilCGO96} could be used to enable the efficient updates.



\balance
\section{Acknowledgments}
This work was partially supported by the EU projects iManageCancer (H2020, \#643529), Bounce (H2020, \#777167) \& NESTOR (Marie Curie \#748945).

\bibliographystyle{abbrv}
\bibliography{vldb_sample}  

\end{document}